\documentclass[reprint,
superscriptaddress,
amsmath,amssymb,
aps,
prb,
]{revtex4-2}
\usepackage[version=3]{mhchem}
\usepackage{graphicx}
\usepackage{bm}
\usepackage{xcolor}

\begin{document}

\preprint{APS/123-QED}

\title{Entropy-driven electron density and effective model Hamiltonian for boron systems}

\author{Chang-Chun He}
\email{scuthecc@scut.edu.cn}
\affiliation{School of Physics and Optoelectronics, South China University of Technology, Guangzhou 510640, China}

\author{Shao-Gang Xu}
\affiliation{Department of Physics, Southern University of Science and Technology, Shenzhen 518055, China}
\affiliation{Quantum Science Center of Guangdong-Hong Kong-Macao Greater Bay Area (Guangdong), Shenzhen 518045, People's Republic of China}

\author{Yu-Jun Zhao}
\affiliation{School of Physics and Optoelectronics, South China University of Technology, Guangzhou 510640, China}

\author{Hu Xu}
\affiliation{Department of Physics, Southern University of Science and Technology, Shenzhen 518055, China}

\author{Xiao-Bao Yang}
\affiliation{School of Physics and Optoelectronics, South China University of Technology, Guangzhou 510640, China}

\date{\today}

\begin{abstract}
The unique electron deficiency of boron makes it challenging to determine the stable structures, leading to a wide variety of forms. In this work, we introduce a statistical model based on grand canonical ensemble theory that incorporates the octet rule to determine electron density in boron systems. This parameter-free model, referred to as the bonding free energy (BFE) model, aligns well with first-principles calculations and accurately predicts total energies. For borane clusters, the model successfully predicts isomer energies, hydrogen diffusion pathways, and optimal charge quantity for \( closo \)-boranes. In all-boron clusters, the absence of B-H bond constraints enables increased electron delocalization and flexibility. The BFE model systematically explains the geometric structures and chemical bonding in boron clusters, revealing variations in electron density that clarify their structural diversity. For borophene, the BFE model predicts that hexagonal vacancy distributions are influenced by bonding entropy, with uniform electron density enhancing stability. Notably, our model predicts borophenes with a vacancy concentration of $\frac{1}{6}$ to exhibit increased stability with long-range periodicity. Therefore, the BFE model serves as a practical criterion for structure prediction, providing essential insights into the stability and physical properties of boron-based systems.

\end{abstract}

\maketitle

\section{Introduction}

Effective models are commonly used to capture the properties of condensed matter systems by focusing on critical interactions. These models often serve as alternatives or preparatory steps for costly experiments   \cite{doi:10.1126/science.1158009}. In realistic materials, model parameters are usually derived from experiments or first-principles calculations, like density functional theory (DFT)   \cite{doi:10.1021/acs.jctc.8b01176,Shao2023}. Electron density, which describes the spatial probability distribution of electrons, is a fundamental quantity that provides key insights into material properties. The Hohenberg-Kohn theorems   \cite{PhysRev.136.B864} establish that the ground state electron density determines the lowest energy of a system with a bijective map. A major objective now is to develop methods for determining effective Hamiltonians and electron densities of complex systems with lower computational costs.

\begin{figure}[ht]
    \centering
    \includegraphics[width=\linewidth]{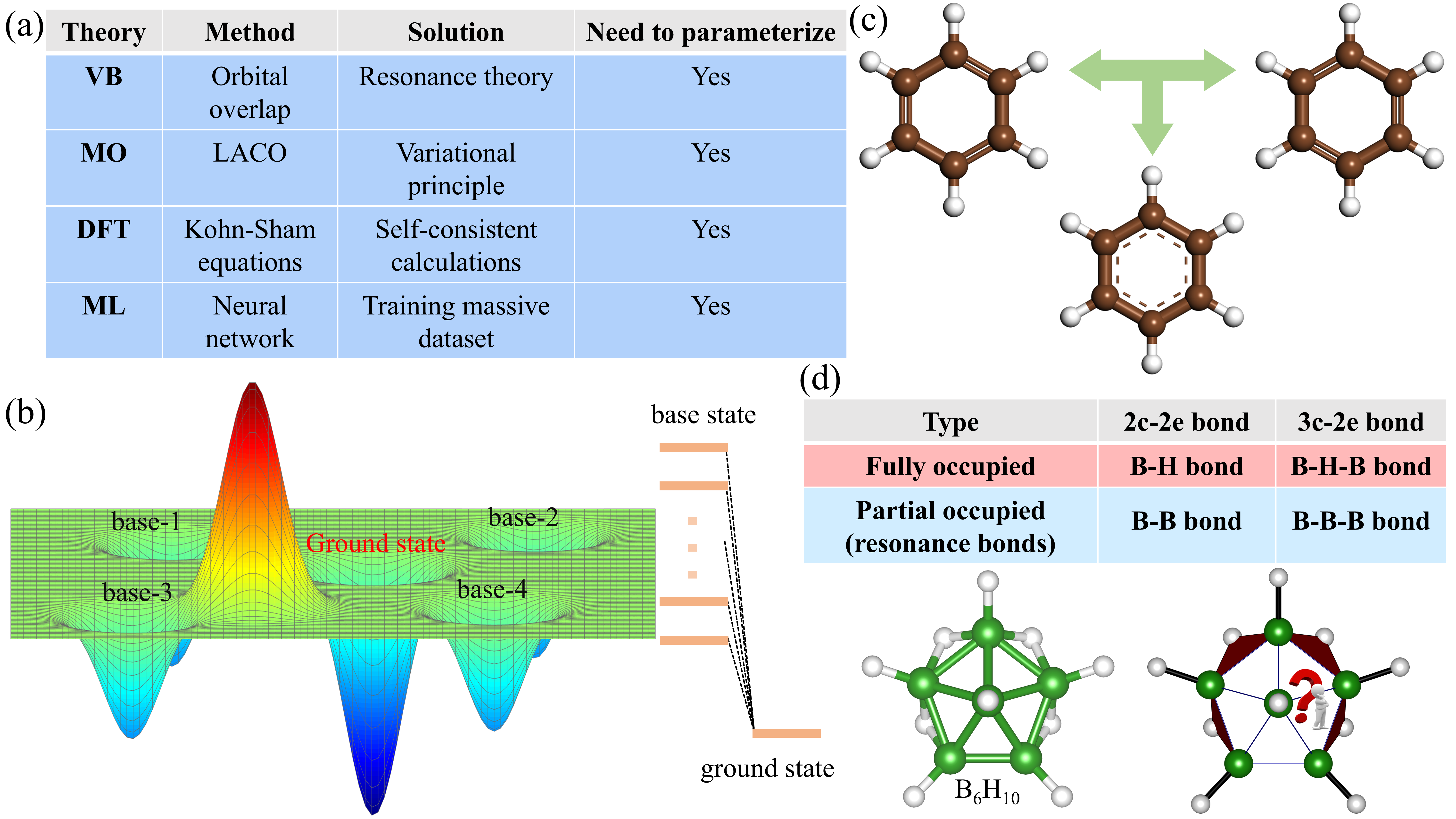}
    \caption{(a) Current popular theories of understanding chemical bonding. (b) A potential energy surface, where red represents high energy, and blue represents low energy. (c) Benzene with two based Kekul$\acute{\textrm{e}}$  structures and the averaged structure. (d) Existing chemical bonds of the boron system and \ce{B5H9} molecule.} 
    \label{fig1}
\end{figure}

Without relying on quantum theory, the shared electron-pair bonding model proposed by Gilbert Lewis   \cite{doi:10.1021/ja02261a002} captures many aspects of chemical bonding, offering a practical framework for understanding molecular structure. To build on the electron-pair model, Langmuir  \cite{doi:10.1021/ja01447a011}  and Kossel  \cite{https://doi.org/10.1002/andp.19163540302} introduced the octet rule, which states that main-group elements in the second period tend to gain, lose, or share electrons to achieve a complete octet in their outermost energy level. However, the octet rule is a phenomenological theory that mainly applies to simple molecules and does not address the complexities of systems with electron delocalization   \cite{QR9571100121}. Valence bond (VB) theory and molecular orbital (MO) theory, through parameterization, offer quantitative estimates of molecular stability   \cite{doi:10.1021/ja00719a006}. As shown in Figure \ref{fig1}(a), VB theory emphasizes the combination of localized atomic orbitals to form chemical bonds, while MO theory describes how delocalized electrons stabilize a molecule by distributing over the entire structure. Developed from self-consistent iterations, DFT uses Kohn-Sham equations   \cite{PhysRev.136.B864} to determine electron density in quantum systems. Although machine learning (ML) algorithms with robust neural networks have become popular, their complex parameters   \cite{doi:10.1021/acs.chemrev.0c01111, Fedik2022} can obscure underlying physical principles, making it challenging to obtain a clear understanding of bonding and stability. Therefore, a parameter-free model based on clear chemical concepts is urgently needed to explore the intrinsic nature of chemical stability.

While the octet rule provides a foundation for understanding electronic structures in molecular systems, it often falls short in explaining electron delocalization. On a potential energy surface (PES) like  Figure \ref{fig1}(b), the ground state is a resonance hybrid structure with lower energy, which is the combination of single Lewis structures. The classic example is benzene, where the actual electron density can be successfully expressed as the average of two base  Kekul$\acute{\textrm{e}}$  structures in Figure \ref{fig1}(c), corresponding to the PES shown in Figure \ref{fig1}(b). These examples show that electron density and effective Hamiltonians can sometimes be simplified in systems with electron delocalization.

Boron, with its electron deficiency   \cite{QR9571100121}, forms unique B-H-B bridge bonds, observed in infrared spectra   \cite{doi:10.1021/ja01148a022} and described as three-center two-electron (3c-2e) bonds. W. N. Lipscomb $et~al$ developed the STYX rules to outline electron assignments among B and H atoms in terms of two-center two-electron (2c-2e) and 3c-2e bonds. However, these rules overlook dominant resonance in systems with electron delocalization   \cite{doi:10.1021/jacs.3c10370}, and previous models struggle to balance stability and electron density predictions. The adaptive natural density partition (AdNDP) method  \cite{OSORIO20231} suggests that charge can be divided into multiple two-electron multicenter bonds, but it does not predict stability. The $\sigma$-bond resonance   \cite{Qiu2023_nat_comm} has been proposed to understand flat boron material bonding but lacks predictive power for structural stability. Gaussian approximation potential   \cite{PhysRevLett.120.156001} accurately describes the PES of boron but does predict electron density distribution. Here, we assume full occupation of B-H and B-H-B bonds, as hydrogen atoms follow the duplet rule. Bonds involved in resonance include B-B 2c-2e and B-B-B 3c-2e bonds, shown in the top panel of Figure \ref{fig1}(d). For the $nido$-\ce{B6H10} molecule, six B-H bonds and four B-H-B bonds are fully occupied as marked in red in the right panel of Figure \ref{fig1}(d). The remaining eight electrons fill the B-B and B-B-B bonds in the pyramid, achieving optimal electron density within the constraint of the octet rule. An accurate model should capture this electron density and the model Hamiltonian.

In this work, we introduce a parameter-free statistical model based on the octet rule for boron systems, including neutral and charged boranes, planar clusters, hollow cages, and monolayers with periodic structures. In boranes, B-H bonds and B-H-B bonds are localized, leading to a relatively sparse PES. In all-boron clusters, the PES is dense, with localized bonds at the edges of hexagonal vacancies. The delocalized bonds in boron monolayers lead to stable structures with long-range periodicity. Total energies and electron density distribution predicted by our model are in good agreement with first-principles calculations, providing deeper insights into the physical properties of these structures.

{\color{black}
\section{Statistical Method}
We take a molecule with $N_{\textrm{bond}}$ bonds and $N_{\textrm{ele}}$ electrons as an example to provide a detailed explanation for calculating the free energy of the bonds, where each bond possesses $n_i$ electrons with a corresponding chemical potential, $\mu_i$. The grand canonical partition function is defined by:  
\begin{equation}
  {\cal{Z}} = \sum_{n_1,n_2,...,n_{N_{\textrm{bond}}}} e^{-\sum_{i=1}^{N_{\textrm{bond}}} (n_i\alpha_i  +\beta E_i)},
\end{equation}
where reduced chemical potential $\alpha_i = -\frac{\mu_i}{k_{\textrm{B}} T}$.  All electrons have been allocated to form single B-H bonds and 3c-2e B-H-B bonds, and $\sum_{i=1}^{N_{\textrm{bond}}}n_i=N_{\textrm{ele}}$. Here, we regard $E_i$ for all bonds as equal, which means that the electron can be equally allocated to each bond and is set to zero for convenience.  $N_{\textrm{ele}}$ electrons need to allocate to $N_{\textrm{bond}}$ bonds,  and the degeneracy  for the  combinations  is the multinomial coefficient $C$, which is
\begin{equation}
    C = \tbinom{N_{\textrm{ele}}}{n_1,n_2,...,n_{N_{\textrm{bond}}}}
\end{equation}
This formula is corresponding to the  multinomial formula
\begin{equation}
    (e^{-\alpha_1}+...+e^{-\alpha_{N_{\textrm{bond}}}})^{N_{\textrm{ele}}} = \sum_{n_1,...,n_{N_{\textrm{bond}}}} \tbinom{N_{\textrm{ele}}}{n_1,...,n_{N_{\textrm{bond}}}} e^{-\sum_{i=1}^{N_{\textrm{bond}}}n_i\alpha_i},
\end{equation}
where $\tbinom{N_{\textrm{ele}}}{n_1,n_2,...,n_6} = \frac{N_{\textrm{ele}}!}{n_1!n_2!...n_{N_{\textrm{bond}}}!}$.  Therefore, we can deduce that the grand  canonical partition function is 
\begin{equation}
 {\cal{Z}}   = (e^{-\alpha_1}+e^{-\alpha_2}+...+e^{-\alpha_{N_{\textrm{bond}}}})^{N_{\textrm{ele}}}.
\end{equation}
 We then express the formula for free energy
\begin{equation}
\begin{split}
      F &= -k_{\textrm{B}}T\log{\cal{Z}}- k_{\textrm{B}}T \sum_{i=1}^{N_{\textrm{bond}}}\alpha_i\frac{\partial \log Z}{\partial \alpha_i}  \\
      &= -N_{\textrm{ele}}k_{\textrm{B}}T \left[ \log{\left(\sum_{i=1}^{N_{\textrm{bond}}} e^{-\alpha_i} \right)}+ \frac{\sum_{i=1}^{N_{\textrm{bond}}} \alpha_i e^{-\alpha_i}}{\sum_{i=1}^{N_{\textrm{bond}}} e^{-\alpha_i}}. \right]
\end{split}
\end{equation}
The average  number of electrons in the bonds $i$ is given by
\begin{equation}
    n_i = -\frac{\partial F}{\partial \mu_i} = N_{\textrm{ele}}\frac{e^{-\alpha_i}}{\sum_{i=1}^{N_{\textrm{bond}}} e^{-\alpha_i}}.
\end{equation}
Here, we define the electron probability for bond $i$ as 
\begin{equation}
    p_i = \frac{n_i}{N_{\textrm{ele}}} = \frac{e^{-\alpha_i}}{\sum_{i=1}^{N_{\textrm{bond}}} e^{-\alpha_i}},
\end{equation}
subsequently, the bonding free energy (BFE) can be expressed by:
\begin{equation}
\begin{split}
    F  &=  N_{\textrm{ele}} k_{\textrm{B}}T\sum_{i=1}^{N_{\textrm{bond}}} p_i\log p_i,
    \label{free_energy}
\end{split}
\end{equation}
and the bonding entropy is defined by 
\begin{equation}
\begin{split}
S &= k_{\textrm{B}}\left( \log {\cal{Z}}  -\sum_{i=1}^{N_{\textrm{bond}}}\alpha_i\frac{\partial \log {\cal{Z}}  }{\partial \alpha_i}        \right) \\
 &= -N_{\textrm{ele}}k_{\textrm{B}} \sum_{i=1}^{N_{\textrm{bond}}} p_i\log p_i.
\end{split}
\end{equation}

It is significant to note that the total energy is roughly proportion to the number of electrons $N_{\textrm{ele}}$ in the system,  and the bonding entropy is also roughly proportion to $\log N_{\textrm{ele}}$. We can rewrite the  BFE by  introducing  the equivalent ``temperature.''
\begin{equation}
\begin{split}
    F(p_1,...,p_{N_{\textrm{bond}}})  =   \frac{N_{\textrm{ele}}k_{\textrm{B}}T_0}{\log N_{\textrm{ele}}} \sum_{i=1}^{N_{\textrm{bond}}}p_i\log(p_i),
\end{split}
    \label{free_BE}
\end{equation}
where the equivalent ``temperature'' $T$ is $\frac{T_0}{\log N_{\textrm{ele}}}$,   $T_0$ is the equivalent ``standard temperature'', and $k_{\textrm{B}}T$ functions as the coefficient to ensure that  BFE is an extensive quantity. Under the octet rule constraint for each atom, Eq. \ref{free_BE} determines the  BFE of a system as a function of $p_i$, where the minimal BFE corresponds to the ground state of a system. 
}

\section{Results and Discussions}

\subsection{Determining the electron density by the parameter-free BFE model}

\begin{figure}[ht]
    \centering
    \includegraphics[width=\linewidth]{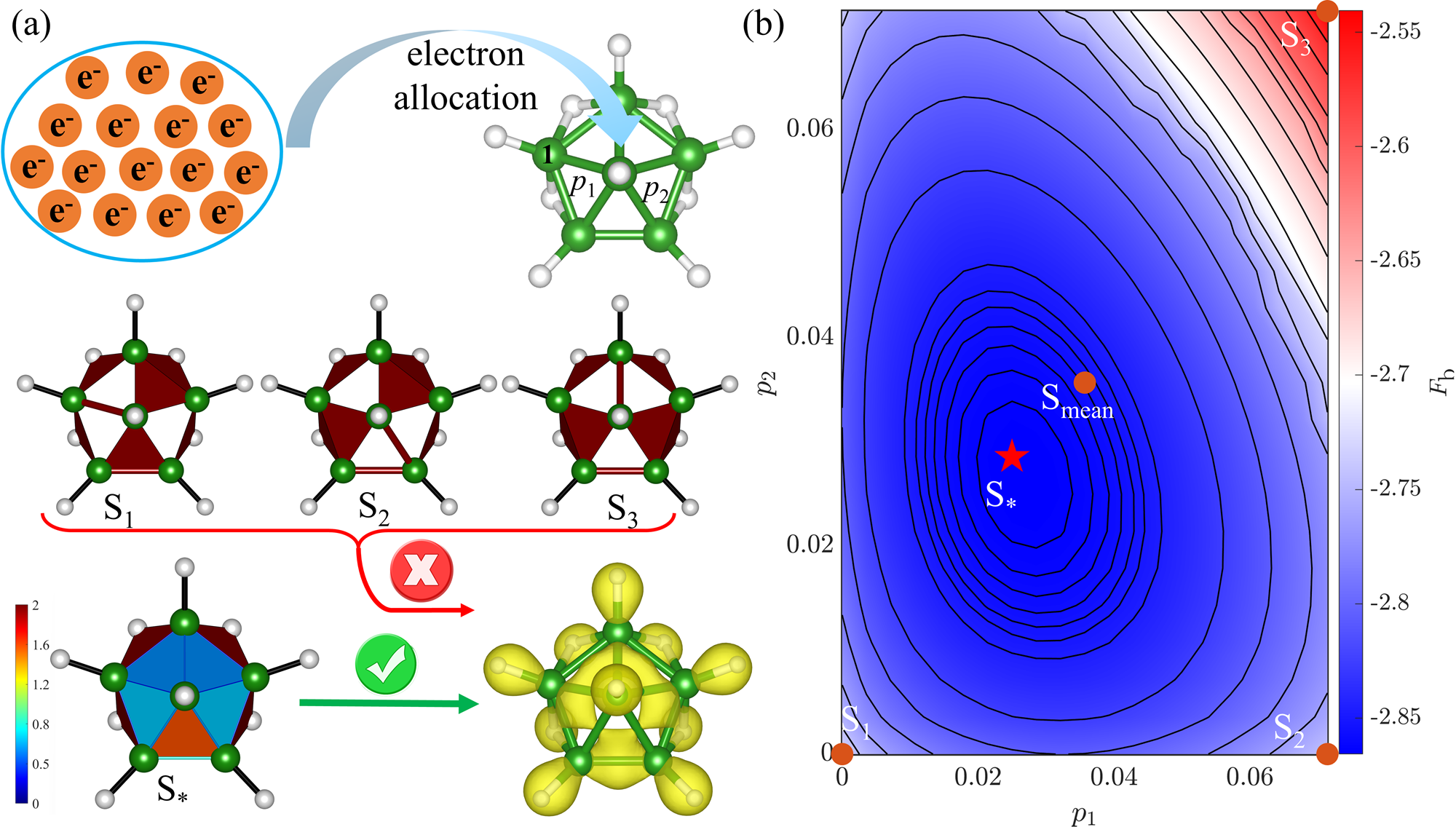}
    \caption{(a) The electron allocation model for borane, three full-occupied chemical bonding of \ce{B5H9}, and the optimal electron density determined by BFE and DFT. (b) The BFE surface varied with $p_1$ and $p_2$.}
    \label{fig2}
\end{figure}

While the proposed 2c-2e and 3c-2e bonds offer some insights into the chemical bonding of boron, quantitatively determining electron density and accurately predicting structural stability remain significant challenges. As illustrated in the upper part of Figure \ref{fig2}(a), the total electrons are distributed across all possible bonds for \ce{B6H10} molecule, necessitating the identification of an optimal electron density distribution. Given that each atom must adhere to the octet rule, a series of resonance structures exist that satisfy local constraints, as depicted in the middle section of Figure \ref{fig2}(a). All bonds are fully occupied or not occupied in these structures, consistent with the STYX rule   \cite{10.1063/1.1740320}. However, these three resonance structures do not accurately represent the actual electron density determined by density functional theory (DFT), as shown in the lower right corner of Figure \ref{fig2}(a), nor do they account for molecular symmetry. Combining the three resonance structures is a way to solve the electron density. However, resonance coefficients are unknown and not easily obtained. Although S$_1$, S$_2$, and S$_3$ comply with the local octet rule, each bond's occupation numbers (ONs) are restricted to either zero or one. This binary representation fails to capture the critical feature of electron delocalization.

Here, we propose a precise yet intuitive statistical model to determine the electron density distribution, balancing the local octet rule for each atom with the global electron delocalization of the total valence electrons. The statistical ensemble averages across all resonance structures, weighted appropriately, can reflect the actual electron density distribution   \cite{PhysRevLett.87.250601, PhysRevLett.122.010601, he2024par}. According to grand canonical ensemble theory, the bonding free energy (BFE) is derived from the grand partition function is referred to Eq. \ref{free_BE}. {\color{black} According to the experimental and theoretical studies  \cite{Piazza2014_nat_com,doi:10.1021/ja203333j}, boron clusters are composed of fragments of triangular lattices with vacancies, while borophenes typically feature the triangle lattice structures with hexagonal vacancies, where the bond lengths in these structures are generally between 1.65 and 1.92 \AA   \cite{doi:10.1021/ja203333j}. For given boron structures, if the distance between two boron atoms is within the bond length range, they are considered to be bonded, forming a two-center, two-electron bond. If the distances between three atoms are all within the bond length range, they are considered to form a three-center two-electron bond. Therefore, the number of all B-B bonds and B-B-B bonds will be easily determined.}

A detailed derivation of the BFE is provided in the supplementary Material (SM)   \cite{supp}. Notably, the BFE exhibits a linear correlation with \(S_{\textrm{b}}\), reaching its minimum when the bonding entropy is maximized by the principle of maximum bonding entropy. 

Each B atom must adhere to the octet rule and each H atom to the duplet rule, as illustrated in the upper right corner of Figure \ref{fig2}(a),  where \( p_1 \) and \( p_2 \) represent the electron probabilities of the two B-B-B bonds. Under this local constraint, the BFE varies with \( p_1 \) and \( p_2 \) as depicted in Figure \ref{fig2}(b). The base resonance structures S$_1$, S$_2$, and S$_3$ are indicated, with corresponding high BFEs marked at the endpoints of the  BFE in Figure \ref{fig2}(b). These points signify that the full occupation of the chemical bonds is unstable due to the lack of electron delocalization. As the mean electron density of the base resonance structures, the S$_{\textrm{mean}}$ has lower BFE than the three resonance structures. The global minimum BFE, corresponding to the electron density distribution S$_*$, is exemplified in the bottom left corner of Figure \ref{fig2}(a), which is lower than S$_{\textrm{mean}}$. S$_*$ supplies a reasonable method to determine the optimal resonance coefficients of the resonance structures as detailed in SM   \cite{supp}. For clarity, colors represent the expected electron count for each bond, ranging from 0 to 2, as shown by the left color bar. The green B-B bond indicates one occupied electron. In contrast, the blue B-B-B bond signifies half an occupied electron, reflecting the resonance structure of all potential Lewis structures and illustrating how electron delocalization enhances stability significantly. S$_*$ aligns well with the electron density calculated by DFT, as shown in the lower section of Figure \ref{fig2}(a). Beyond the typical \ce{B6H10} molecule, the electron densities of other $nido-$B$_n$H$_{n+4}$ molecules are detailed in the Figure S1 of SI   \cite{supp}, demonstrating strong consistency with DFT calculations. This evidence suggests that the BFE accurately describes electron density, indicating a tendency for electrons to exhibit a uniform distribution within the constraints of the octet rule.

From the view of information entropy, the bonding entropy serves as a quantitative measure of how uninformative a probability distribution is, ranging from zero (completely informative) to $\log(p)$  (completely uninformative). Given the available information, we adopt the most uninformative distribution possible by selecting the distribution that maximizes bonding entropy. Opting for a distribution with lower entropy would imply the assumption of information that we do not possess. Thus, the distribution with maximum entropy emerges as the most reasonable choice, forcing the electron distribution to be more uniform, achieving the lowest electrostatic potential state. The method inspires us to establish a realistic electron density distribution for delocalized electron systems with the octet rule and the maximum entropy principle.

\subsection{The neutral and charged boranes}

\begin{figure}[ht]
    \centering
    \includegraphics[width=\linewidth]{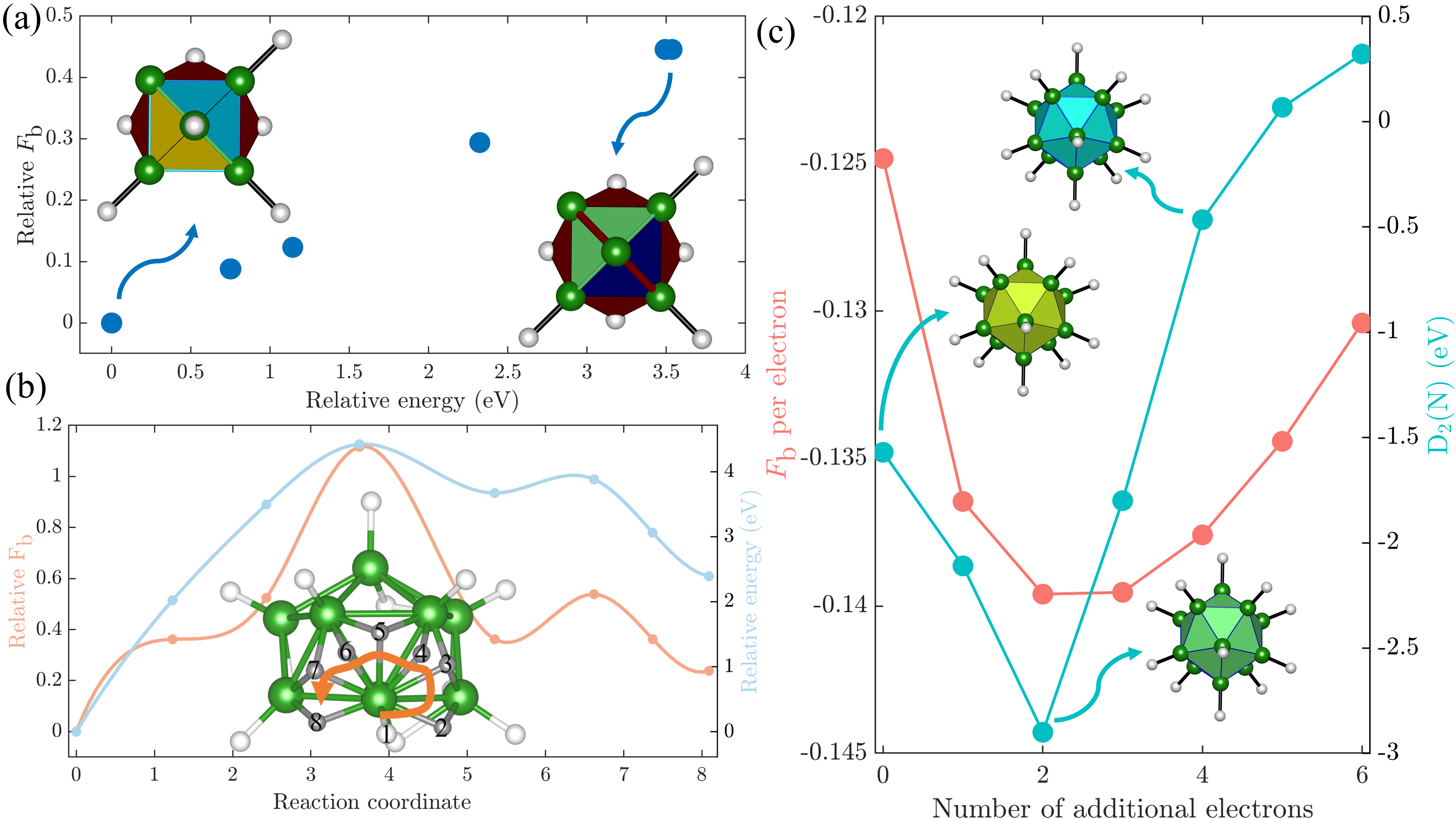}
    \caption{(a) The energy scatter plot of \ce{B5H7} isomers predicted by BFE and DFT. (b) BFE and DFT determine the transition energy path calculation. (c) The total energy of \ce{B12H12} molecule for the number of additional electrons.}
    \label{fig3}
\end{figure}

Since the minimal BFE has accurately determined electron density, the BFE can be treated as a model Hamiltonian to describe the total energy of the boron system, thereby predicting structural stability. We use the \ce{B5H7} molecules as an example, generated by removing two hydrogen atoms from the \ce{B5H9} molecule, as illustrated in Figure \ref{fig2}(a), to construct a series of borane isomers. Each isomer can be analyzed using the BFE model, with the electron density determined by the minimal BFE.   As shown in Figure \ref{fig3}(a), the relative energies of the isomers correlate well with the predicted \( F_{\textrm{b}} \) calculated by the BFE model, indicating that our model effectively discriminates the stability of \ce{B5H7} isomers. Among these, the most stable isomer loses one B-H bond and one B-H-B bond, as shown in the left part of Figure \ref{fig3}(a). The color of each bond reflects the localization and delocalization according to the color bar in Figure \ref{fig2}(a), with the electron density in this stable structure appearing notably more uniform compared to the higher-energy structure on the right.

For the \ce{B8H12} molecule, we construct a diffusion path for the hydrogen H atom along the direction indicated by the arrows in Figure \ref{fig3}(b). Along this path, the H atom traverses the top site of one B atom, the bridge site between two B atoms, and the center site formed by three B atoms. The transition state images along the diffusion path can be described using the BFE model, enabling us to infer the variation of \( F_{\textrm{b}} \) along the path, represented by the orange line in Figure \ref{fig3}(b). To validate our predictions, we employed the CI-NEB method   \cite{Henkelman2000} to accurately calculate the minimal energy path, as shown by the light blue line in Figure \ref{fig3}(b). The energy variation trend of the H atom during the diffusion process aligns well with the predictions of the BFE model. Notably, both methods indicate that the structure with the H atom at the bridge site is typically higher in energy than that with the H atom at the center site of three B atoms. This difference arises because the H atom at the bridge site forms an additional B-H-B 3c-2e bond. In contrast, the H atom at the center site forms three B-H-B 3c-$\frac{2}{3}$e bonds, with $\frac{2}{3}$e contributing to a more substantial electron delocalization effect that enhances structural stability. Notably, the transition state corresponding to the maximum energy barrier is identified as index 4, where the H atom is located at the bridge site, as detailed in Figure \ref{fig3}(b).

For the \( closo \)-borane B\(_n\)H\(_n\) clusters, the most prominent and well-known species is the \( closo \)-dodecaborate \(\ce{B12H12^{2-}}\) dianion   \cite{doi:10.1021/ja01501a076,doi:10.1021/ic50012a002}. Its derivatives are also significant in various fundamental and applied research fields   \cite{doi:10.1021/cr00010a005,https://doi.org/10.1002/anie.201702237}, with boron-based neutron capture therapy for cancer being particularly notable   \cite{doi:10.1021/cr980493e}. Interestingly, two extra electrons are often required to enhance structural stability, reflecting the electron-deficient nature of boron. Previous theoretical studies have employed the coupled-cluster singles and doubles (CCSD) approach with the cc-pVDZ basis set to elucidate the reasons for the high stability of the dianion icosahedron   \cite{doi:10.1021/acs.jpca.1c09167}. The second difference of the total energies, defined as \( D_2(N) = 2E_{N} - E_{N-1} - E_{N+1} \), where \( N \) represents the charge of the cluster, provide valuable insights for determining the most stable configuration. As shown by the green line in Figure \ref{fig3}(c), \( D_2(2) \) is the lowest among various charge states, indicating that the addition of two electrons yields the most stable state for the \(\ce{B12H12}\) molecule. Herein, the BFE model also predicts that these additional two electrons significantly enhance structural stability. The neutral \(\ce{B12H12}\) molecule has 48 electrons, and the BFE model determines the optimal electron density distribution when \( F_{\textrm{b}} \) is minimized, as illustrated in Figure \ref{fig3}(c). If one additional electron is added, resulting in 49 electrons, the allocation aims for a new optimal uniform electron density. However, the \( F_{\textrm{b}} \) averaged per electron is lower than that of the neutral state, suggesting that the extra electron compensates for boron's electron deficiency. Although boron is inherently electron-deficient, an accumulation of excess electrons can raise the system's electrostatic potential energy, causing structural instability. Thus, it is imperative to achieve a balance. The BFE model further predicts that the most stable configuration should accommodate two additional electrons, demonstrating the model's strong predictive capability. Furthermore, the other  \( closo \)-borane B\(_n\)H\(_n\) ($n=6-11$) clusters are analysed in Figure S2 and S3 in SI   \cite{supp}, also in  agreement with the DFT calculations   \cite{doi:10.1021/ic980359c}.

\subsection{The planar clusters,  hollow cages, and bilayer clusters}

\begin{figure*}[ht]
    \centering
    \includegraphics[width=\linewidth]{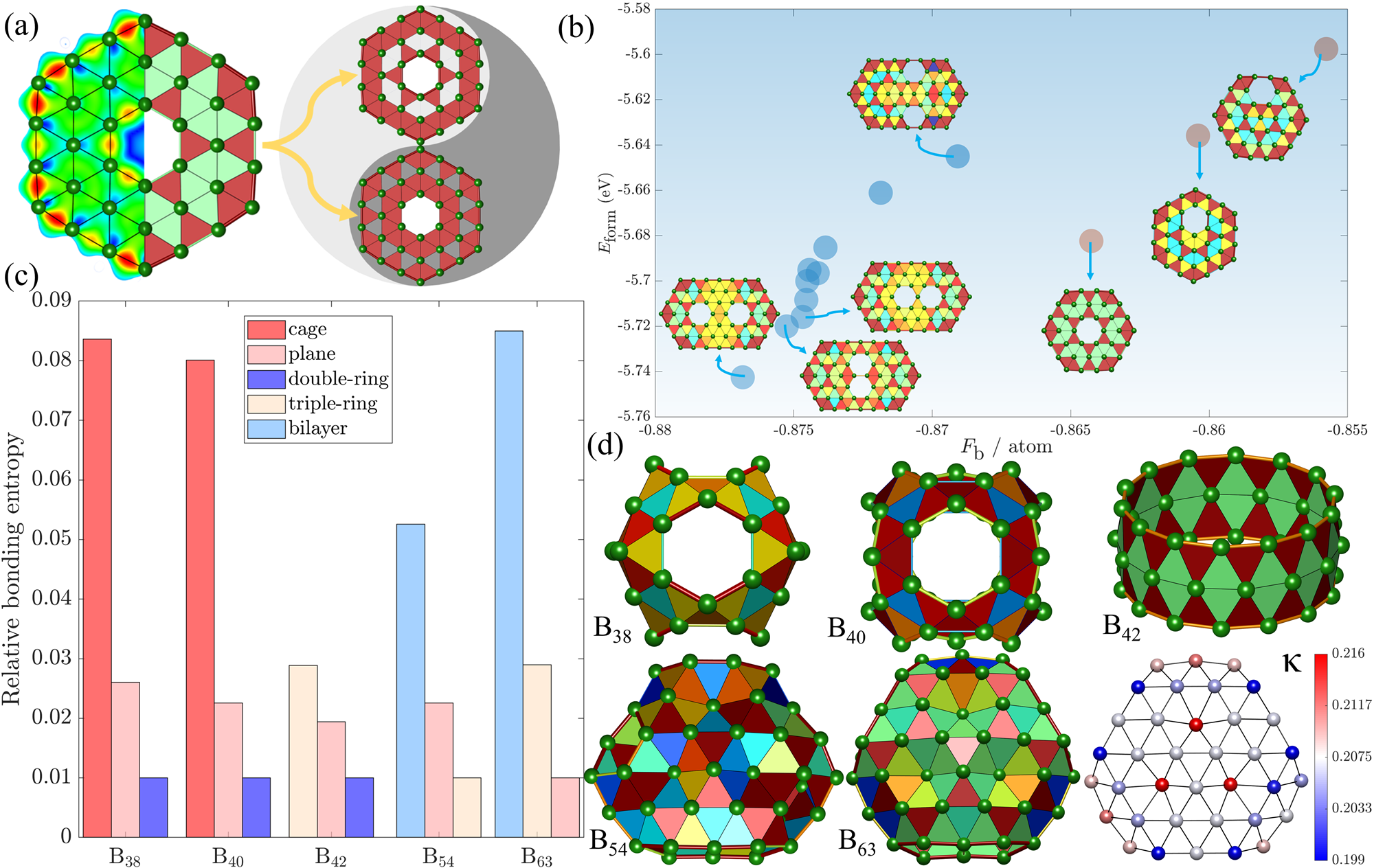}
    \caption{(a) The electron density of \ce{B36} and the decomposed  Kekul$\acute{\textrm{e}}$  structures. (b) The energy predicted by the BFE model and DFT for \ce{B36} with one vacancy and \ce{B56} with two vacancies. (c) The  evolution prediction of  different styles of boron clusters, with the electron density and local aromaticity plotted in (d).}
    \label{fig4}
\end{figure*}

The evolution of geometric structures and chemical bonding for boron clusters is significant in searching for new-generation boron-based materials. However, achieving this understanding has proven challenging, necessitating over a decade of sustained collaborative experimental and theoretical investigations. The highly stable planar \(\ce{B36}\) cluster, which exhibits six-fold symmetry and possesses a central hexagonal hole, was observed experimentally, as shown in the inset of Figure \ref{fig4}(a)   \cite{Piazza2014}. The valence shell of the \(\ce{B36}\) cluster is expected to be filled with eight shared electrons according to the octet rule, based on 2c-2e and 3c-2e bonds. Here, the 2c-2e bond refers to the B-B bond, while three nearest neighbor B atoms form the 3c-2e bond. Assuming there are \(n_1\) 2c-2e bonds and \(n_2\) 3c-2e bonds in the \(\ce{B36}\) cluster, we have the equations \(2n_1 + 2n_2 = 36 \times 3\) and \(4n_1 + 6n_2 = 36 \times 8\), leading to \(n_1 = N_{\textrm{ele}}\) and \(n_2 = 36\). The electron density determined by the BFE model, shown in the inset of Figure \ref{fig4}(a), is consistent with DFT calculations. Notably, the 3c-2e bonds adjacent to the edge B atoms are fully occupied, while other 3c-2e bonds are half-occupied. Similarly, the 2c-2e bonds adjacent to the B atoms at the six external vertices are fully occupied, whereas those adjacent to the six internal vertices and located at the center of the external edge are half-occupied, exhibiting high consistency with the DFT calculations.

Notably, all B-B bonds vanish except those at the edges, as 3c-2e bonds replace the inner B-B bonds. Given that the coordination number of inner B atoms is six, 3c-2e bonds are preferentially formed to ensure that each B atom satisfies the octet rule, reflecting the electron-deficient nature of boron. Each external edge B atom shares two fully occupied B-B single bonds. In comparison, internal edge B atoms share two half-occupied B-B single bonds due to their greater coordination and the prevalence of 3c-2e bonds. The half-occupied state can be viewed as a neutralization between full and empty occupancy as illustrated in Figure \ref{fig4}(a), and the "Taijitu" representation suggests that two fully occupied electronic densities achieve an optimal balanced distribution through complementary interactions. Similar to the benzene molecule, the electron density in the \(\ce{B36}\) cluster can be regarded as the average of two H{\ "u}ckel structures.

In addition to electron density analysis, we can utilize the \(F_{\textrm{b}}\) to predict the structural stability of B clusters. For the \(\ce{B36}\) clusters, the hexagonal hole can interchange with neighboring B atoms, as illustrated in the right part of Figure \ref{fig4}(b). Notably, the \(\ce{B36}\) cluster with six-fold symmetry exhibits the lowest \(F_{\textrm{b}}\) and formation energy, while the energy of other \(\ce{B36}\) clusters increases with decreasing \(F_{\textrm{b}}\), maintaining a strong linear relationship.  As the energy of the three \(\ce{B36}\) clusters gradually increases, their electron density distributions become increasingly non-uniform, with a growing number of fully occupied B-B 2c-2e bonds. In the case of the \(\ce{B56}\) cluster, characterized by a double-hexagonal vacancy, varying vacancy distributions induce diverse electron densities and structural stabilities. \(F_{\textrm{b}}\) exhibits a linear correlation with \(E_{\textrm{form}}\), indicating that the BFE model effectively predicts stability for clusters with double-hexagonal vacancies. Furthermore, these specific double holes are characteristic of two-dimensional stable B sheets, such as the \(\alpha\)-B sheet   \cite{PhysRevLett.99.115501,PhysRevB.77.041402,Wu2012_acs_nano}, serving as a key motif for stable B sheets.

For the larger B clusters, we will demonstrate how $F_{\textrm{b}}$ can provide insights into the evolutionary process of B clusters with varying sizes using several boron clusters with similar total energies. For each boron cluster, the bar chart from left to right corresponds to isomers with gradually increasing energy as shown in Figure \ref{fig4}(c), and the lowest relative bonding entropy is set to 0.01 for easy presentation, in which the most stable B clusters are plotted in Figure \ref{fig4}(d), and other clusters are shown in SI   \cite{supp}. For the \ce{B38}   \cite{C4NR01846J} and \ce{B40}   \cite{Zhai2014}, the most stable structures are both cage-like with hexagonal vacancies of the largest bonding entropy, where the planar clusters and double-ring clusters are less stable with lowest bonding entropy. Since the number of B atoms with a coordinate number (CN) of 5  in the cage clusters is larger than that of planar and double-ring clusters, as well as the number of 3c-2e bonds (see Figure \ref{fig4}(d)),  the ONs of cage clusters are more uniform, inducing the largest bonding entropy. When the size becomes larger, the most stable cluster evolves to triple-ring (\ce{B42}) and double-layer (\ce{B54}   \cite{https://doi.org/10.1002/ejic.202000473} and \ce{B63}   \cite{doi:10.1021/acs.jpclett.4c00566}). At the same time, other types of boron clusters are relatively more energetic with less bonding entropy.

Particularly, we take \ce{B63} as an example to define the local aromaticity of B atom $i$,  $\kappa_i=-\sum_{j} p_{ij}\log{p_{ij}}$, where $j$ traverses all bonds connect with B atom $i$. The six central inward-buckled B atoms with red color display the largest value, exhibiting typical aromaticity, while the edge atoms with blue color show less aromaticity,   which is also in agreement with the Nucleus-independent chemical shifts (NICS) value as referred to SI   \cite{supp}. The six hexagons centered around these central inward-buckled B atoms resemble benzene rings (\ce{C6H6}), and the local solid aromaticity effectively enhances the system's stability.

\subsection{Borophenes with long-periodic structures}

\begin{figure}[ht]
    \centering
    \includegraphics[width=\linewidth]{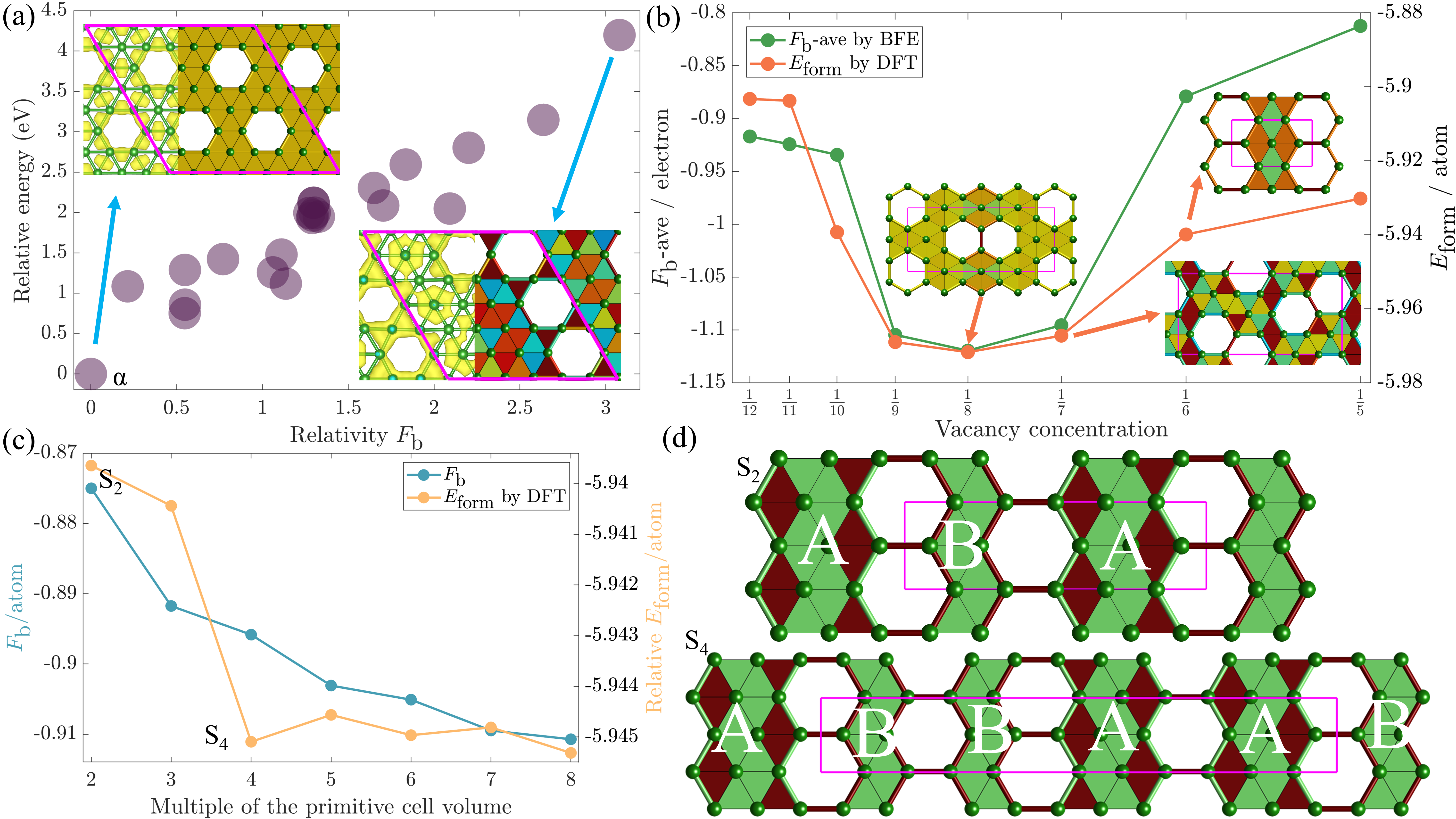}
    \caption{(a) The energy scatter plot of borophene with various vacancy distributions predicted by BFE and DFT. (b) The energy scatter plot  of different vacancy concentrations predicted by BFE and DFT.  {\color{black} (c) The $E_{\textrm{form}}$  predicted by DFT and $F_{\textrm{b}}$ predicted by the BFE model for borophene with $\eta$ = 1/6. (d) The most stable borophene structures S2/S4 in a two/four-period arrangement with $\eta$ = 1/6, along with the electron density determined by the BFE model.}}
    \label{fig5}
\end{figure}

For boron monolayers,  the triangular sheets by carving hexagonal vacancies will be stable, where vacancy concentration $\eta$ is defined by the ratio of the number of hexagon vacancies to the number of atoms in the pristine triangular sheet   \cite{Wu2012_acs_nano}. Previous works reported that the $\alpha$-borophene is the most stable monolayer for the largest cohesive energy per atom in DFT calculation   \cite{PhysRevLett.99.115501, PhysRevB.77.041402} for the vacancy concentration is $\eta=\frac{1}{9}$,  which is formed by removing one B atom from a $3\times3$ supercell of triangular lattice. To verify the model capability, we generate the no-equivalent structures with four hexagonal vacancies in $6\times6$ supercell using  SAGAR code developed by our group   \cite{HE2021110386} and calculate the $F_{\textrm{b}}$ by BFE model and total energies by DFT calculations of these structures, exhibiting the excellent consistency as shown in Figure \ref{fig5}(a). Among these structures, the most stable structure is $\alpha$-borophene, and the most unstable one is in the bottom right of Figure \ref{fig5}(a), which contains four    X-type  B atoms with CN of 4. Note that the borophene with  X-type B atoms is unstable for the absence of resonance   \cite{Qiu2023_nat_comm}, and the quantitative BFE model also shows that electron density distribution is pretty non-uniform,  contributing to the instability. Two-dimensional borophene exhibits polymorphism,  where the ground state structures are intrinsically independent of $\eta$. The octet rule only applies to the borophene with $\eta=\frac{1}{9}$. In contrast, borophenes with other vacancy concentrations can not obey the octet rule due to the electron deficiency   \cite{C8NR01230J}, in which the additional electron should be compensated into the borophene. Therefore,  the average $F_{\textrm{b}-ave}$ which is defined by the ratio of $F_{\textrm{b}}$ to total electrons to evaluate the structural stability for various vacancy concentrations as detailed in SI   \cite{supp}. As demonstrated in Figure \ref{fig5}(b), the prediction by the BFE model has good correlations with the DFT calculations. Moreover,  the borophene with $\eta=\frac{1}{8}$ is the most energetic rather than $\alpha$-borophene among these borophenes with $\eta=\frac{1}{12}-\frac{1}{5}$, as an evidence to show the generality of our model. The borophene with  $\eta=\frac{1}{8}$,  which has the lowest $F_{\textrm{b}}$-ave,  can be viewed as the adjoint of two mirrored $\alpha$-borophene   \cite{PhysRevMaterials.5.044003}, implying the polymorphism of two-dimensional boron. The electron density distributions of borophene with other vacancy concentrations are detailed in SI   \cite{supp}, where the electron density will aggregate around the vacancies, diminishing the structural stability.

{\color{black}
Although it seems that there is a significant difference between the DFT and BFE models at high vacancy concentrations in borophene, the BFE model can accurately describe the relationship between vacancy concentration and structural stability, and it can reliably predict the most stable borophene structure at the same concentration. The large discrepancy at high concentrations could be due to the introduction of the concept of compensating charge. As the vacancy concentration increases, the compensating charge becomes larger, which may lead to a greater difference between the results of the BFE model and DFT calculations.
}

Exceptionally, we choose the borophene with $\eta=\frac{1}{6}$ to investigate the polymorphism of borophene further. We start from the borophene with  $\eta=\frac{1}{6}$ plotted in Figure \ref{fig5}(b) as the primitive cell and construct the supercell with the same vacancy concentration along the horizontal direction to investigate whether there are long-periodic structures with lower $F_{\textrm{b}}$-ave. As shown in Figure \ref{fig5}(c), we find that the $F_{\textrm{b}}$-ave and $E_{\textrm{form}}$  gradually decrease with the multiple of primitive cell, uncovering that the long-periodic borophenes are often more stable than the short-periodic borophene. The DFT calculations have also confirmed the prediction that the four- and eight-time supercells are more stable than the unit cell. In Figure \ref{fig5}(d), the S2  can be viewed as being composed of  A and B types of nanoribbons assembled, which will further increase the bonding entropy and decrease the $F_{\textrm{b}}$-ave, contributing the structural stability. S4,   viewed as A-A-B-B type,  exhibits a higher degree of disorder, which increases the bonding entropy and explains boron's polymorphic nature.

{\color{black} Although the BFE model failed to accurately predict that the energy of S4 in Figure 5(c) is the lowest, it is still able to identify the S4 structure as having the lowest energy within the same supercell. Thus, for a given cell with specific vacancy concentrations, the BFE model provides the correct ranking trend for isomers of boron structures, which significantly enhances the screening efficiency.}

\section{Conclusions}
We propose a parameter-free statistical model for describing boron systems based on grand canonical ensemble theory, combining the octet rule. The bonding free energy (BFE) model has perfectly described the electron density and the total energies of boron systems. For the borane clusters, the B-H bonds and B-H-B bonds are localized with full occupation, and other delocalized B-B bonds are determined by the BFE model, which has successfully predicted the isomer energies of borane clusters as well as the hydrogen diffusion energy pathway. For $closo$-$\textrm{B}_n\textrm{H}_n$ molecule, our model accurately reveals that adding two extra electrons maximizes stability, demonstrating broad applicability. Without localized B-H and B-H-B bonds, all-boron clusters have greater degrees of freedom,  exhibiting stronger delocalization. The BFE model provides a systematic understanding of the geometric structures and chemical bonding of size-selected boron clusters,  which is crucial for discovering new boron-based nanostructures. The evolution as a function of size can be described by the BFE model, which is ascribed to the electron density distribution, implying the polymorphism of boron clusters. For borophene, periodic boundary condition induces stronger electron delocalization and polymorphism. The distribution of hexagonal vacancies is determined by bonding entropy, with uniform electron density contributing to structural stability. The origin of borophene's polymorphism is linked to the reduction of bonding entropy through electron compensation. In particular, we demonstrate that borophene with a vacancy concentration of $\frac{1}{6}$ exhibits higher structural stability, often associated with long-range periodicity. Therefore, the BFE model can serve as a criterion for structure prediction, providing deeper insights into the physical nature of these structures.

%

\end{document}